\RequirePackage{fix-cm}
\documentclass[twocolumn]{svjour3}          
\smartqed  

\usepackage{lineno,hyperref}
\usepackage{amssymb}
\usepackage{graphicx}
\usepackage[ruled,vlined,linesnumbered]{algorithm2e}
\usepackage{float}
\usepackage{color}
\usepackage{soul}
\usepackage{amsmath}
\usepackage{caption}
\usepackage{makecell}
\usepackage{enumitem} 
\usepackage{makecell}
\usepackage{subcaption}
\usepackage{multirow,siunitx}
\begin{document}

\title{E-MIIM: An Ensemble Learning based Context-Aware Mobile Telephony Model for Intelligent Interruption Management}


\titlerunning{An Ensemble Learning based Context-Aware Mobile Telephony}        

\author{Iqbal H. Sarker$^{1,2}$, A.S.M. Kayes$^{3}$, Md Hasan Furhad$^{4}$, Mohammad Mainul Islam$^{5}$ and Md Shohidul Islam$^{6}$}

\authorrunning{Sarker et al.} 

\institute{$^1$ Chittagong University of Engineering and Technology, Bangladesh. $^2$ Swinburne University of Technology, Melbourne, VIC-3122, Australia. Corresponding Email: msarker@swin.edu.au \\ $^3$ La Trobe University, Melbourne,  VIC 3086, Australia. \\ $^4$ Canberra Institute of Technology, ACT, Australia. \\  $^5$ Extend View Inc., Los Angeles, CA, USA. \\  $^6$ University of California, Riverside, CA, USA.}

\date{Received: date / Accepted: date}

\maketitle

\newcommand\blfootnote[1]{%
	\begingroup
	\renewcommand\thefootnote{}\footnote{#1}%
	\addtocounter{footnote}{-1}%
	\endgroup
}

\makeatletter
\def\footnoterule{\relax%
	\kern 0pt
	\hbox to \columnwidth{\hfill\vrule width .95\linewidth height 0.6pt\hfill}
	\kern .1pt}
\makeatother

\blfootnote{This paper has been published in the Journal of ``AI \& SOCIETY", Springer Nature.}

\begin{abstract}
Nowadays, mobile telephony interruptions in our daily life activities are common because of the inappropriate \textit{ringing notifications} of incoming phone calls in different contexts. Such interruptions may impact on the work attention not only for the mobile phone owners but also the surrounding people. \textit{Decision tree} is the most popular machine learning classification technique that is used in existing context-aware mobile intelligent interruption management (MIIM) model to overcome such issues. However, a single decision tree based context-aware model may cause over-fitting problem and thus decrease the \textit{prediction accuracy} of the inferred model. Therefore, in this paper, we propose an \textit{ensemble machine learning} based context-aware mobile telephony model for the purpose of intelligent interruption management by taking into account multi-dimensional contexts and name it ``E-MIIM''. The experimental results on individuals' real life mobile telephony datasets show that our E-MIIM model is more effective and outperforms existing MIIM model for predicting and managing individual's mobile telephony interruptions based on their relevant contextual information.

\end{abstract}

\section{Introduction}
Nowadays, mobile phones became an important part of our daily life. The increasing quality and quantity of cell phones has made them the most personal and essential communication devices today. According to \cite{sarker2018MobileDataScience}, users' interest on mobile phones is growing day-by-day and the amount of mobile cellular subscriptions is nearly the amount of individuals on the earth, and it reaches even 100\% of the population in the developed countries. The growing adoption and the increased popularity of mobile phones have dramatically modified the approach to act and communicate like mobile telephony with people in the current real world. These mobile phones are thought to be `always on, continually connected' devices \cite{chang2015investigating}. However, the mobile phone users aren't continually attentive and conscious of incoming mobile telephony communications owing to their daily things in their real life activities. For this reason, generally mobile telephony occurs interruptions by the incoming phone calls that not only produce disturbance for the device owners but also for the people nearby as well. Such mobile telephony interruptions could produce embarrassing situation not only in an official environment, e.g., professional meeting, but also additionally have an effect on different major activities like lecturing in a class, examining patients by a doctor in a medical hospital or driving a vehicle etc. Generally, these types of mobile telephony interruptions could increase errors and stress in a working environment like corporate office, medical center, software industry etc. and thus decrease overall employee performance \cite{pejovic2014interruptme}.

Typically, we expect a phone call to ring to inform the users concerning the incoming calls from another. However, ringing phone at an inconvenient moment are often causes interruptions to the present task or social scenario \cite{sarker2018Unavailability}. According to the Basex BusinessEdge report \cite{spira2005cost}, the mobile telephony interruptions greatly impact on employee productivity. They report that mobile telephony interruptions consume 28\% of the knowledge worker's day, that is predicated on surveys and interviews conducted by Basex over the eighteen months amount encompassing high-level employee, senior executives at the end-user organizations, and executives at corporations that manufacture cooperative business information tools. This interprets into twenty eight billion lost man-hours once a year to corporations within United States alone. It leads to a loss of \$700 billion, considering a mean remuneration of \$25/hour for a knowledge employee, according to Bureau of Labor Statistics \cite{laborStatistics}. In another study associated with the execution time of primary tasks, Bailey et al. \cite{bailey2006need} have shown that once interrupted users need from 3\% to 27\% longer to complete the tasks, and commit double the quantity of errors across the tasks. Thus, intelligently managing mobile telephony interruptions in our real-world life, has become one of the key research areas
in the domain of IoT (Internet of Things) and smart mobile phone services. Therefore, in this work, we aim to build a \textit{machine learning} based context-aware mobile telephony model that can predict individual's mobile telephony activities and manage such interruptions intelligently utilizing their own mobile data.

Mobile phones automatically record individuals' mobile telephony activity data and corresponding contextual information associated with their owners in the device's log \cite{sarker2016phone}. Their ability to log such activities offers the potential to analyze the telephony behavior of people. Modeling user mobile telephony behaviors such as individual's responses of incoming phone calls, analyze different behavioral patterns, and eventually predict the longer term call response behaviors from phone log data, can be used for building data-driven interruption management system that shows intelligent assistance for the end mobile phone users in their daily activities \cite{sarker2018research}. Let's consider a context-aware mobile telephony interruption management example. Say, Alice, a smart phone user, works in a software company, her workplace, as a programmer analyst. She attends a daily regular meeting in the morning between 09:00AM and 10:00AM at her office to discuss about the updates of ongoing projects with her team members. Typically, she declines the incoming phone calls in that time period as she doesn't want to be interrupted with phone calls throughout the meeting. Because the interruptions might not solely disturb herself, additionally could disturb her team members within the meeting. However, if the phone call comes from the managing director of that company, she needs to answer the phone call because it probably to be necessary for her by taking into account the importance, although she is in a professional meeting. Thus, the mobile telephony behavior of a user could be changed depending on her contextual situation. Hence, [decline, answer] are the user mobile telephony activities and [in the morning, at office, at meeting, managing director] are the associated contextual information. An intelligent mobile telephony behavior model utilizing her phone log data could predict Alice's telephony behavior supported her contexts, in which she answers or declines the incoming phone calls in her daily work routine. Such telephony model utilizing Alice's phone log data, can be used to build an intelligent context-aware interruption management system for the user Alice, so that she can enjoy context-aware intelligent services for minimizing mobile telephony interruptions in her real world life.

Machine learning classification techniques are more popular and common to build such context-aware intelligent prediction model \cite{sarker2019classifications}. Among the classification techniques, decision tree \cite{quinlan1993} is a well-known technique that can be used to model and predict user telephony behavior utilizing individual's phone log data consisting of their mobile telephony activities and corresponding contextual information. An example of contexts are - temporal context that represents particular day-of-the week (e.g., Monday) and time-of-the day (e.g., in the morning) information, spatial context that represents user's location (e.g., at office), social context that represents users' social situation (e.g., in the meeting) or social relationship (e.g., colleague in office), which are relevant to individual's mobile telephony activities according to the given example above \cite{sarker2018BehavMiner}. In the area of context-aware mobile services, a number of researchers \cite{hong2009context} \cite{lee2007deploying} \cite{zulkernain2010mobile} \cite{sarker2019machine} have used decision tree classification technique to model mobile phone usage behavior for various purposes. In particular, Zulkernain et al. \cite{zulkernain2010mobile} have proposed a decision tree based intelligent system, known as ``MIIM'' (Mobile Intelligent Interruption Management) that intelligently handles mobile interruptions according to users' current contexts. However, the classification rules generated by the decision tree have low reliability in prediction \cite{sarker2019recencyminer}.
According to \cite{freitas2000understanding}, a decision tree cannot ensure that a discovered classification rule will have a high predictive accuracy due to \textit{over-fitting problem}, i.e., it performs well on the training dataset but getting relatively poor performance to make predictions on unseen test cases. The reason is that a single decision tree based context-aware model may be biased in some cases to capture the behavior properly because of considering the precedence of contexts for the entire dataset. Thus, the \textit{challenge} is to minimize the biasness of contexts and to improve the prediction accuracy of a context-aware mobile telephony model in order to build an intelligent interruption management system. 

In this paper, we address this issue and propose an \textit{ensemble learning} based context-aware mobile telephony model for intelligent interruption management, and name it ``E-MIIM''. In our model, we first extract the contextual features from the training dataset and pre-process the contexts to fit for an ensemble machine learning based context-aware modeling. As an ensemble learner, we use random forest technique that generates multiple decision trees by taking into account the subsets of data, rather than a single decision tree for the entire dataset, to achieve our goal. Random forest ensemble learner consists of a number of decision trees and outputs the majority vote of individual decision trees. Once the ensemble learning based context-aware mobile telephony model has been built, the prediction results are computed for the unseen test cases. The effectiveness of E-MIIM model presented in this paper over existing MIIM model is examined by doing experiments on the real mobile telephony datasets consisting of individual's diverse mobile telephony activities and corresponding contextual information.

The rest of the paper is organized as follows. Section \ref{Related Work} reviews the related work. In section \ref{Contextual Information}, we discuss about mobile telephony interruptions and the relevant contextual information. In Section \ref{Methodology}, we present our ensemble learning based context-aware model for predicting and managing mobile telephony interruptions. We report the experimental results in Section \ref{Evaluation}. Some key observations of our model are summarized in Section \ref{Discussion} and finally Section \ref{Conclusion} concludes this paper and highlights the future work.

\section{Related Work}
\label{Related Work}
A significant amount of research has been done on mobile interruptions, particularly telephony interruptions by voice calls, their management systems and usefulness of these systems for the users. For instance, Khalil et al. \cite{khalil2006context} have shown the usefulness of interruption management system by conducting a user survey. In their survey, they investigate the contextual information relevant to the context-aware telephony. In their work, they enhance the agreement between callers and callee with the aim of minimizing mobile interruptions. They found the low availability rate, around 53\% of the time, of the users to answer the incoming phone calls. In another work, Toninelli et al. \cite{toninelli2009s} have reported a review of movement based reaction to approaching calls of different users and demonstrate that maximum users would prefer not to be interrupted while in a professional meeting or working in a group or outside of the working environment like driving a vehicle or sleeping and ignore the incoming telephony in these circumstances.

To minimize such interruptions, Pejovic et al. \cite{pejovic2014interruptme}, build an interruption management library dependent on various sensors of Android phones. In \cite{smith2014ringlearn}, the authors present a new approach to smartphone interruptions that maintains the quality of mitigation under concept drift with long-term usability. In their method, they use online machine learning and accumulates names for intrude on causing occasions, e.g., incoming calls, utilizing verifiable experience examining without requiring additional intellectual load for the user's sake. Another exploration is unobtrusively extraordinary dependent on user interface. Authors present a multiplex user interface for dealing with incoming approaches of cell phones \cite{bohmer2014interrupted}. This plan arrangement handles the issue that calls can hinder simultaneous application employments. They expanded the choices for taking care of incoming phone calls and presented contemplations for potential outcomes to delay calls and multiplex the call warning with the simultaneous application.

A number of authors have studied on call interruption management systems based on individual's calendar events, e.g., meeting. For instance, in a calendar based study, Khalil et al. \cite{khalil2005improving} state that calendar entries are a good cue as to whether a person is available or unavailable for a phone call. They utilize calendar data to consequently arrange mobile phones as needs be to oversee interruptions. The calendar allows the user to define specific tasks or events with duration, temporal domain and other attributes \cite{sarker2016evidence}. Salovaara et al. \cite{salovaara2011phone} have conducted a study and show that 31\% of the incoming phone calls were unavailability related, e.g., the users are unavailable to answer the phone calls for various reasons such as meetings, lectures, appointments, driving, sleeping. To minimize interruptions, Dekel et al. \cite{dekel2009minimizing} design an application based on special keywords such as `lectures', `meetings', `appointments'. In \cite{zulkernain2010context} and \cite{zulkernain2010mobile}, the authors utilize the calendar information in order to design context-aware interruption management system. Seo et al. \cite{seo2011pyp} utilize user's schedule to define policy rules in their context-aware phone configuration management application to improve cell phone awareness. In these approaches, the interruption handling rules are based on the static temporal segments according to their scheduled appointments in their individual's calendar information, e.g., the user is unable to answer the incoming call whenever s/he is in a scheduled event of the calendar (e.g., a meeting between 13:00 and 14:00). 

To build an effective mobile telephony behavior for minimizing interruptions for individual users, the calendar based approaches might not be effective in their real world life. In many cases of our daily life, the phone call is not disruptive even though the user is engaged in an ongoing task or social situation and the call is welcome as it provides a needed mental break from the current task \cite{de2007should}. According to \cite{grandhi2010technology}, 24\% of cell phone users feel the need to answer a phone call when they are in a meeting. Rosenthal et al. \cite{rosenthal2011using} have shown through users survey that 35\% of the participants want to receive phone calls at work, while other participants do not want. Sarker at al. \cite{sarker2016evidence} have shown that the existence of an event of a particular time segment in the calendar is inadequate to infer the actual behavior of individuals for their various calendar events. According to \cite{lovett2010calendar}, the calendar does not provide the consistently accurate representation of the real world due to events not occurring or the events may occur outside their allotted time window in the calendar. The main drawback of these systems is that the approaches are not data-driven and the contextual information or rules used by the applications are not automatically discovered; users need to define and maintain the rules manually. In general, users may not have the time, inclination or expertise to do this. 

In contrast, individual's mobile telephony activities and corresponding contextual data recorded in the device logs, can be used as a rich resource for analyzing their telephony behavior \cite{sarker2016phone}. Machine learning techniques can play a vital role to build the relevant context-aware prediction model utilizing such phone log data. For instance, a number of authors \cite{hong2009context} \cite{lee2007deploying} \cite{zulkernain2010mobile} \cite{sarker2019machine} have used decision tree machine learning classifier to model mobile phone usage behavior utilizing such mobile phone data. In particular, an intelligent context-aware interruption management system (MIIM) using machine learning technique has designed by Zulkernain \cite{zulkernain2010mobile}. In their approach, they use decision tree machine learning classifier to build their prediction model for the purpose of managing mobile interruptions. However, the classification rules generated by the single decision tree cannot ensure that a discovered classification rule will have a high predictive accuracy in the resultant context-aware model. As a result, it may decrease the prediction accuracy due to over-fitting problem while designing the tree, i.e., it performs well on the training dataset but getting relatively poor performance to make predictions on unseen test cases.

Unlike the above discussed approaches, in this work, we present an \textit{ensemble learning} based context-aware model for better prediction accuracy in managing mobile interruptions.

\section{Mobile Telephony Interruptions and Contexts}
\label{Contextual Information}

Let, $Act = \{Act_1,Act_2, ..., Act_n\}$  be a set of mobile telephony activities of an individual user $U$, each action $Act_i$ represents a particular phone call activity for that user. In the real world, the common mobile telephony activities of an individual mobile phone user are \cite{sarker2018Unavailability} - (i) answering the incoming phone call by the user for a particular time period or duration, i.e., `Accept', (ii) instantly decline the incoming phone call by the user, i.e., `Reject', (iii) the phone rings for an incoming call but the user misses the call, i.e., `Missed', and (iv) making a phone call to a particular person, i.e., `Outgoing'. Except the outgoing call, all the mobile telephony activities are related to incoming communications from another person and might have a chance to cause interruptions as the users are not always attentive and responsive to incoming phone calls in their daily activities in the real world.

As mentioned earlier, smart mobile phones automatically record each phone call activity of individuals and corresponding contextual information in device's log. These are temporal context, spatial context, or social context that are relevant to individual's mobile telephony activities, highlighted in Introduction with an example. As in this paper, we aim to improve the existing context-aware model `Mobile Intelligent Interruption Management (MIIM)' system presented by Zulkernain et al. \cite{zulkernain2010mobile} in terms of prediction accuracy for unseen context-aware test cases, we highlight the relevant contextual information associated with the models. For instance, as the temporal context, they use the day of week along with time of day such as morning, evening etc. In addition to the temporal context, they also use location and schedule as contextual information in their system. Unlike the characteristics of these contexts, we use data-driven contextual information in our context-aware mobile telephony model `E-MIIM' for the purpose of minimizing mobile interruptions. For instance, as temporal contexts, we use time-series historical data recorded in the phone log of individuals, which vary from user-to-user. In addition to the temporal context, we also use the spatial context and social context that might have also an influence on individual mobile phone users to make phone call decisions \cite{sarker2017designing}. For instance, an individual's phone call response behavior at her `office' may be well different from her response when she is at `home', which represents an example of spatial context. Similarly, individual's interpersonal social relationship such as mom, friend, colleague, boss, significant one, or unknown \cite{sarker2018DataCentricSocialContext} may have strong influence on individuals to handle call response decision in the real world. For example, a user typically `declines' an incoming phone call during an event official meeting, however, she `answers' if the incoming call comes from her `boss'. Thus, in our context-aware mobile telephony model, we take into account all these contextual information relevant, in order to predict the unseen test cases for the purpose of building intelligent mobile interruption management system for the benefit of end mobile phone users.

\section{Methodology: E-MIIM}
\label{Methodology}
In this section, we present our ensemble learning based context-aware model in order to predict mobile user telephony behavior based on multi-dimensional contexts utilizing individual's mobile phone data.

\subsection{Dataset and Contextual Features Extraction}
Real-world smartphone data usually comprise a set of features whose interpretation depends on some \textit{contextual information} such as temporal, spatial, or social context discussed in the earlier section. Such contextual features and their patterns with mobile telephony are of high interest to be discovered in order to achieve our goal. For this purpose, in this paper, we have used a real world dataset, known as Reality Mining dataset collected by Massachusetts Institute of Technology (MIT) for their Reality Mining project~\cite{eagle2006reality}. This dataset contains individuals mobile telephony activities over nine months period and corresponding contextual information. In order to build our context-aware model, first we extract the contextual information that have an influence on making a phone call decision according to the usages patterns of individuals. For example, time-series data can be used as a temporal context. However, the raw time-series dataset of an individual user need to be preprocessed to convert into nominal values. 

To use such time-series temporal information in modeling users' mobile telephony behavior, we use our earlier behavior-oriented time-series segmentation technique BOTS \cite{sarker2017individualized} for converting into nominal values (e.g., time segments). This technique generates a number of segments having similar behavioral characteristics of individuals. In addition to temporal context, user location and individual's unique phone number can be used as spatial context, and social context respectively. An example of spatial contexts are home, office, market, MIT, Harvard etc that exist in the dataset. Additionally, we generate data-centric social context using individual's unique mobile phone number in the dataset, which can be used as a social relationship between individuals \cite{sarker2018DataCentricSocialContext}. The reason for generating such data-centric social context is to effectively build personalized model as it is able to differentiate person-to-person, even in a similar relationship like `friend'. For instance, one's mobile telephony behavior may vary between `best friend' and `close friend', even though both persons represent same relationship category `friend'. Our generated data-centric social relational context resolves this issue which is important for personalization. We have also preprocessed the mobile telephony activity classes using call duration and call type \cite{sarker2016behavior}, such as Accept, Reject, and Missed, that are related to incoming communications and interruptions discussed earlier and represented as Class 1, Class 2, and Class 3 respectively. Therefore, we predict these classes according to the current contexts of individual users in our ensemble learning based context-aware telephony model for the purpose of building an intelligent interruption management system to minimize the mobile interruptions.

\subsection{Contextual Ensemble Learning}
Once the dataset is ready to process, we use random forest ensemble learning which is one of the most popular and powerful machine learning algorithms, in order to build our context-aware model. An ensemble method is a technique that combines the predictions from multiple machine learning algorithms together to make more accurate predictions than any individual model. A random forest \cite{breiman2001random} is an ensemble classifier consisting of many decision trees by taking into account the subsets of data, rather than a single decision tree for the entire dataset, where the final predicted class for a test example is obtained by combining the predictions of all individual trees. Random forests combine bootstrap aggregation (bagging) \cite{breiman1996bagging} and random feature selection \cite{amit1997shape} to construct a collection of decision trees exhibiting controlled variation.

To generate the random forest, we take into account all the relevant contexts discussed above. The generated random forest consists a number of decision trees that can be used a separate classifier like a single decision tree based model. At each node in a tree, $d \ll D$ features are randomly selected from the D available features in the dataset, and the node is partitioned according to the Gini Index \cite{breiman1984lclassification}. The Gini Index is calculated by subtracting the sum of the squared probabilities of each class from one. For a binary split, the Gini Index of a node n can be expressed as - 

\begin{equation}
$$I_G(n) = 1 - \sum_{i=1}^{c}{p_i}^2$$
\end{equation}

where $p_i$ is the probability of an object being classified to a particular class. The best possible binary split is the one which maximizes the improvement in the Gini index.

\begin{equation}
$$\Delta I_G(n_p) = I_G(n_p) - p_l I_G(n_l) - p_rI_G(n_r)$$
\end{equation}

where $p_l$ and $p_r$ are the proportions of examples in node $n_p$ that are assigned to child nodes $n_l$ and $n_r$ respectively.

Finally, we combine the trees to form a single, strong learner by taking into account the majority vote, in order to build an effective behavior model for individual mobile phone users, for the purpose of increasing the prediction accuracy for a particular context or contextual association. For instance, we generate $N$ random decision trees utilizing the training dataset. The terminal nodes of each decision tree represents the behavior classes and the edges are associated with the corresponding contexts that are used for similarity matching in prediction. For a particular test case, each random decision tree generated in the model may predict different outcomes according to the contexts in the tree. In order to make the final prediction result for that contexts, we calculate and store all the predicted output for each tree and perform the majority voting among the $N$ trees. After building the model, we evaluate the user behavior model using the testing dataset and measure the effectiveness in terms of prediction accuracy.

\section{Evaluation and Experimental Results}
\label{Evaluation}	
To evaluate the effectiveness of our context-aware model E-MIIM, we have conducted experiments on the real mobile phone datasets for predicting individual's telephony behavior. For this purpose, we have done our experiments on ten phone log datasets for different individual mobile phone users. We also present an experimental evaluation comparing our context-aware model E-MIIM with the existing MIIM model in terms of prediction accuracy for unseen test cases. In the following subsections, we define the evaluation metrics that are taken into account in our experiment and present the experimental results and discussion.

\subsection{Evaluation Metric}
In order to measure the prediction accuracy, we compare the predicted response with the actual response (i.e., the ground truth) and compute the accuracy in terms of:

\begin{itemize}
	\item Precision: ratio between the number of phone call behaviors that are correctly predicted and the total number of behaviors that are predicted (both correctly and incorrectly). If TP and FP denote true positives (number of examples predicted positive that are actually positive) and false positives (number of examples predicted positive that are actually negative) then the formal definition of precision is \cite{witten1999weka}:
	
	\begin{equation}
	Precision = \frac{TP}{TP + FP}
	\end{equation}
	
	\item Recall: ratio between the number of phone call behaviors that are correctly predicted and the total number of behaviors that are relevant. If TP and FN denote true positives (number of examples predicted positive that are actually positive) and false negatives (number of examples predicted negative that are actually positive) then the formal definition of recall is \cite{witten1999weka}:
	
	\begin{equation}
	Recall = \frac{TP}{TP + FN}
	\end{equation}
	
	\item F-measure: a measure that combines precision and recall is the harmonic mean of precision and recall. The formal definition of F-measure is \cite{witten1999weka}:
	
	\begin{equation}
	Fmeasure = 2 * \frac{Precision * Recall}{Precision + Recall}
	\end{equation}
	
	\item Kappa: is a metric that compares an observed accuracy with an expected accuracy (random chance). The formal definition of Kappa is \cite{witten1999weka}:
	
	\begin{equation}
	Kappa = \frac{(observed \; accuracy - expected \; accuracy)}{(1 - expected \; accuracy)}
	\end{equation}
\end{itemize}

\begin{table*}[htbp!]
	\centering
	\caption{The prediction results for each individual class of our E-MIIM model for user U1}
	\label{results-ensemble-01}
	\begin{tabular}{|c|c|c|c|c|c|} 
		\hline
		\bf Class & \bf TP Rate & \bf FP Rate & \bf Precision & \bf Recall & \bf F-measure \\  
		\hline
		Class 1 & 0.710 & 0.052 & 0.724 & 0.710 & 0.712 \\ 
		\hline
		Class 2 & 0.782 & 0.006 & 0.887 & 0.782 & 0.831 \\ 
		\hline
		Class 3 & 0.794 &  0.036 & 0.826 & 0.794 & 0.810 \\ 
		\hline
	\end{tabular}
\end{table*}

\begin{table*}[htbp!]
	\centering
	\caption{The prediction results for each individual class of our E-MIIM model for user U2}
	\label{results-ensemble-02}
	\begin{tabular}{|c|c|c|c|c|c|} 
		\hline
		\bf Class & \bf TP Rate & \bf FP Rate & \bf Precision & \bf Recall & \bf F-measure \\  
		\hline
		Class 1 & 0.808 & 0.069 & 0.825 & 0.808 & 0.817 \\ 
		\hline
		Class 2 & 0.651 & 0.002 & 0.932 & 0.651 & 0.766 \\  
		\hline
		Class 3 & 0.602 & 0.024 & 0.707 & 0.602 & 0.650 \\ 
		\hline
	\end{tabular}
\end{table*}

\begin{figure*}[htbp!]
	\centering
	\begin{subfigure}[b]{.49\textwidth}
		\includegraphics[width=.9\textwidth]{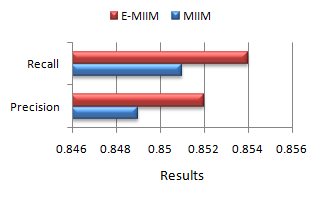}
		\caption{Comparing prediction results for a random user.}
		\label{fig:comparison-D1}
	\end{subfigure}
	\begin{subfigure}[b]{.49\textwidth}
		\includegraphics[width=.9\textwidth]{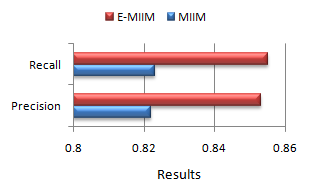}
		\caption{Comparing prediction results for another random user.}
		\label{fig:comparison-D2}
	\end{subfigure}
	\caption{Comparing prediction results of our context-aware model E-MIIM with existing model MIIM in terms of weighted precision and recall for two different random users.}
	\label{fig:effectiveness-individuals}
\end{figure*}

\begin{figure*}[htbp!]
	\centering
	\includegraphics[width=.5\linewidth, keepaspectratio]{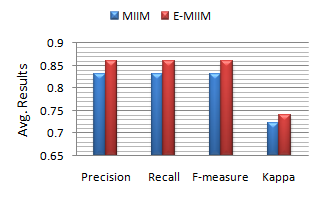}
	\caption{Comparing average prediction results (utilizing all the datasets) of our context-aware model E-MIIM with existing model MIIM in terms of precision, recall, f-measure, and kappa value.}
	\label{fig:comparison-all}
\end{figure*}

\subsection{Experimental Results and Analysis}
To evaluate our E-MIIM model in context-aware test cases, we employ the most popular 10-fold cross validation technique on each dataset. In general, cross validation technique estimates how accurately a predictive model will perform in practice. In a prediction problem, a model is usually given a dataset of known data on which training is run (training dataset), and a dataset of unknown data (or first seen data) against which the model is tested. The 10-fold cross validation breaks data into 10 sets of size N/10. It trains the model on 9 sets and tests it using the remaining one set. This repeats 10 times and we take a mean accuracy rate. To show the effectiveness of each model, we compute and compare the accuracy, in terms of precision, recall, f-measure and kappa statistics, defined above.

\subsubsection{Prediction Results of E-MIIM}
In this experiment, we show the prediction results of our context-aware model E-MIIM. For this, Table \ref{results-ensemble-01} and Table \ref{results-ensemble-02} show the prediction results in terms of TP (True Positive) rate, FP (False Positive) rate, Precision, Recall, and F-measure, for each individual telephony behavior class (Class 1, Class 2, Class 3) utilizing the datasets of two different individual users. If we observe Table \ref{results-ensemble-01} and Table \ref{results-ensemble-02}, we see that for each class the FP rate (instances falsely classified as a given class) is very low, and the values of TP rate, Precision, Recall, and F-measure are significantly high, which estimate the better prediction accuracy. Thus, the overall experimental results in Table \ref{results-ensemble-01} and Table \ref{results-ensemble-02} show that our ensemble learning based context-aware model E-MIIM is able to effectively predict each telephony behavior class of individual users according to their usage patterns in the dataset.

\subsubsection{Effectiveness Comparison}
In this experiment, we compute and compare the effectiveness in terms of precision, recall, f-measure, and kappa of our context-aware model E-MIIM, with existing mobile intelligent interruption management model MIIM \cite{zulkernain2010mobile}. For each model, we utilize the same datasets in order to compare the models fairly. As our model is personalized, we first show the comparing results for individual users selected randomly. To show the effectiveness for individual users, Figure \ref{fig:effectiveness-individuals} shows the relative comparison of precision, recall, f-measure and kappa for two different individuals utilizing their own datasets. For each prediction model, we use the same training and testing set of data based on 10-fold cross validation technique, for the purpose of fair evaluation and comparison. If we observe Figure \ref{fig:effectiveness-individuals}, we see that our ensemble learning based context-aware model E-MIIM gives better prediction results in terms of precision and recall than existing MIIM model for each individual user's dataset.

In addition to individual's comparison, we also compute the prediction results for all the datasets of ten individual users mentioned above and show the relative comparison of avg. (average) results of these datasets. For this, we calculate the average precision, recall, f-measure and kappa value of all ten datasets. The average results are shown in Figure \ref{fig:comparison-all}. If we observe Figure \ref{fig:comparison-all}, we find that our context-aware model E-MIIM consistently outperforms existing model MIIM for different datasets of individuals. In particular, our model gives the higher prediction results in terms of average precision, recall, f-measure and kappa statistic. The reason for getting better result is that we generate multiple decision trees with subset of data and take the average result for the final outcome. Thus, it reduces the variance through averaging over the single decision tree learners, and the randomized stages decreasing correlation between distinctive learners in our model. As a result, the model becomes more effective than existing prediction model, when applying on mobile phone data consisting of multi-dimensional contexts and corresponding users' telephony activities. 

Finally, according to the experimental results shown in Figure \ref{fig:effectiveness-individuals} and Figure \ref{fig:comparison-all}, we can conclude that our ensemble learning based context-aware model E-MIIM is more effective than existing model MIIM for modeling and predicting individual's context-aware telephony activities for the purpose of managing mobile interruptions.

\section{Discussion}
\label{Discussion}
Overall, our context-aware model E-MIIM utilizing individual's telephony data is fully individualized and behavior-oriented. Compared to the existing model MIIM, the prediction accuracy in terms of precision, recall, f-measure is improved when our ensemble learning based approach is used, as shown in Figure \ref{fig:effectiveness-individuals}, and Figure \ref{fig:comparison-all}. Although, this model requires to generate a number of decision trees to output a prediction result for a particular contextual information, it is effective in terms of prediction accuracy for unseen test cases. The following are a few key discoveries from our machine learning based study for mobile intelligent interruption management. 

\begin{itemize}
	\item To predict individuals' telephony behavior for a particular context, ensemble learning based model, e.g., random forest model generating multiple decision trees based on relevant contexts is more effective than a single decision tree based model. In our experiments, we have shown the effectiveness in terms of precision, recall, f-measure and kappa values for different users utilizing their own mobile phone datasets consisting of their telephony activities and corresponding contextual information.

	\item We have observed a significantly lower prediction accuracy of existing interruption management model MIIM, compared to our context-aware model E-MIIM. The reason is that existing model MIIM cannot capture the behavior patterns properly in multi-dimensional contexts as it takes into account only one decision tree that may cause over-fitting problem. Consequently, these approaches have low prediction accuracy compare to our ensemble learning based context-aware telephony behavioral model.
	
	\item Our model does not depend on the static number of contexts for modeling individuals' telephony behavior. For more contexts, it will perform better than existing MIIM model, as we take into account generating multiple decision trees to produce an output. In this work, we mainly focus on the ensemble learning based telephony behavior modeling based on a number of contexts relevant to the users. As such, we use temporal, spatial, and social contexts available in the collected mobile phone datasets to show the effectiveness of our model.
	
	Overall, the ensemble learning based context-aware telephony behavioral model is more effective according to its prediction results. Though it's computational complexity is higher because of generating multiple trees than a single decision tree based model, it shows the effectiveness in terms of prediction accuracy. This model is very helpful for large mobile phone datasets as it divides into subsets of data and contexts and perform the local search to generate the corresponding tree from many different starting points.
\end{itemize}

\section{Conclusion and Future Work}
\label{Conclusion}
In this paper we have presented an ensemble learning based context-aware mobile telephony model for intelligent interruption management. Experimental results on individuals' real life mobile telephony datasets show that our mobile telephony model E-MIIM is more effective and outperforms existing MIIM model for predicting and managing individual's mobile telephony interruptions based on their relevant contextual information. No prior knowledge is needed in employing our technique. We believe that our experimental analysis based on machine learning techniques can help both the researchers and application developers on modeling mobile phone user behavior based on multi-dimensional contexts and building corresponding real-life applications in order to provide them better context-aware personalized services according to their own behavior.

Developing an intelligent mobile telephony interruption management system based on this prediction model that handles the incoming phone calls to minimize interruptions can be a future work in order to assess the usability of this model in application level.

\bibliographystyle{plain}
\bibliography{EMIIM}

\end{document}